\documentclass{osa-article}

\journal{osajournal}

\articletype{Research Article}

\usepackage{lineno}

\usepackage{upgreek}
\usepackage{amsmath}
\usepackage{physics}
\usepackage{xcolor, soul}
\sethlcolor{green}
\setcopyright{}

\journal{osajournal}

\begin{document}

\title{Machine learner optimization of optical nanofiber-based dipole traps for cold $^{\textrm{87}}$Rb atoms}

\author{Ratnesh K. Gupta,\authormark{1} Jesse L. Everett,\authormark{1,*}, Aaron D. Tranter \authormark{2}, Ren\'e Henke,\authormark{1} Vandna Gokhroo,\authormark{1} Ping Koy Lam\authormark{2}, and S\'ile {Nic Chormaic}\authormark{1}}

\address{\authormark{1}Light-Matter Interactions for Quantum Technologies Unit, Okinawa Institute of Science and Technology Graduate University, Onna, Okinawa 904-0495, Japan.\\
\authormark{2}Centre for Quantum Computation and Communication Technology, Department of Quantum Science, Research School of Physics, The Australian National University, Canberra, ACT 2601, Australia.}

\email{\authormark{*}jesse.everett@oist.jp}




\begin{abstract}
In two-color optical nanofiber-based dipole traps for cold alkali atoms, the trap efficiency depends on the wavelength and intensity of light in the evanescent field,  and the initial laser-cooling process.  Typically, no more than one atom can be trapped per trapping site.  Improving the trapping efficiency can  increase the number of filled trapping sites, thereby increasing the optical depth.  Here, we report on the implementation of an in-loop stochastic artificial neural network machine learner  to trap $^{87}$Rb atoms in an uncompensated two-color evanescent field dipole trap by optimizing the absorption of a near-resonant, nanofiber-guided, probe beam.  By giving the neural network control of the laser cooling process, we observe an increase in the number of dipole-trapped atoms by $\sim$ 50\%, a small decrease in their average temperature from 150 $\upmu$K to 140 $\upmu$K, and an increase in peak optical depth by 70\%. The machine learner is able to quickly and effectively explore the large parameter space of the laser cooling control to find optimal parameters for loading the  dipole traps.  The increased number of atoms  should facilitate studies of collective atom-light interactions mediated via the evanescent field.
\end{abstract}

\section{Introduction}
Optical nanofiber (ONF) interfaces for controlling cold atom-light interactions have proven to be a very promising platform for hybrid quantum technologies, with several applications demonstrated in recent years \cite{gouraud_demonstration_2015, solano_super-radiance_2017, Ruddell:17, PhysRevLett.122.253603,Ray_2020, rajasree_generation_2020,pucher2021atomic}.  Atoms trapped in a two-color, evanescent field dipole trap around an ONF  \cite{le_kien_atom_2004,vetsch_optical_2010} are shown to have long lifetimes and useful spatial and collective properties \cite{asenjo-garcia_exponential_2017, corzo_waveguide-coupled_2019}, while facilitating direct integration into a fiber network. This further increases their appeal for practical quantum information schemes.  For many applications, trapping the atoms without affecting the energies between the relevant atomic levels is desirable. This can be achieved by tuning the two trapping laser wavelengths to so-called magic values \cite{le_kien_state-insensitive_2005,lacroute_state-insensitive_2012} where the differential ac-Stark shift cancels. Such a compensated trap has been demonstrated for laser-cooled Cs atoms \cite{vetsch_optical_2010, corzo_large_2016}. For Rb atoms, few such wavelengths exist, and compensation can only be achieved  for pure linear or pure circular polarization of the trapping fields, and only for some upper energy levels \cite{arora2012state}. Notably, pure polarization cannot be achieved in the evanescent field of the ONF \cite{PhysRevA.73.053823, PhysRevApplied.11.064041, Joos:19, tkachenko_polarisation_2019, PhysRevResearch.2.033341}.  

Another challenge with using Rb in experiments, is that the tens of MHz light shift broadens the absorption profile of a standard probe beam that is near-resonant to the cooling transition, e.g. the $5S_{1/2}\rightarrow5P_{3/2}$ transition for $^{87}$Rb \cite{lee_inhomogeneous_2015}.  This makes it difficult to determine  the number and temperature of trapped atoms in real-time. Furthermore, optimization of the optical depth, and thus the atom-light interaction, depends on both the number and temperature of the trapped atoms.

With Rb being one of the most widely used atomic species in cold atom quantum experiments, it is worthwhile finding alternative techniques to optimize the loading of ONF-based dipole traps despite the aforementioned obstacles. A complete quantitative description of atom dynamics for the optimal loading of  evanescent field dipole traps from a magneto-optical trap (MOT)  is generally intractable. This is compounded by the difficulty in measuring the atom number in real-time to enable optimization of the trapping sequence during an experiment.  One tractable approach to optimizing physical systems which preclude access to a quantitative description is that of online optimization. This  allows a learner or agent to directly interact with the system in an in-loop setting, thereby providing the physical system with new parameters to implement while receiving feedback on the system's response. Such an approach has already been used to optimize the loading of atoms into a Bose-Einstein condensate using a Gaussian process learner \cite{wigley_becml_2016, davletov_becml_2020}. For larger parameter spaces, it is favorable to leverage methods which are computationally more expensive to minimize calls made to the physical system. One possibility is to use deep learning methods, which have demonstrable success with high dimensional problems such as image classification \cite{schawinski_2017} and regression \cite{wang2018deep}. 

\begin{figure}[htbp]
\centering
\includegraphics[width=8.4cm]{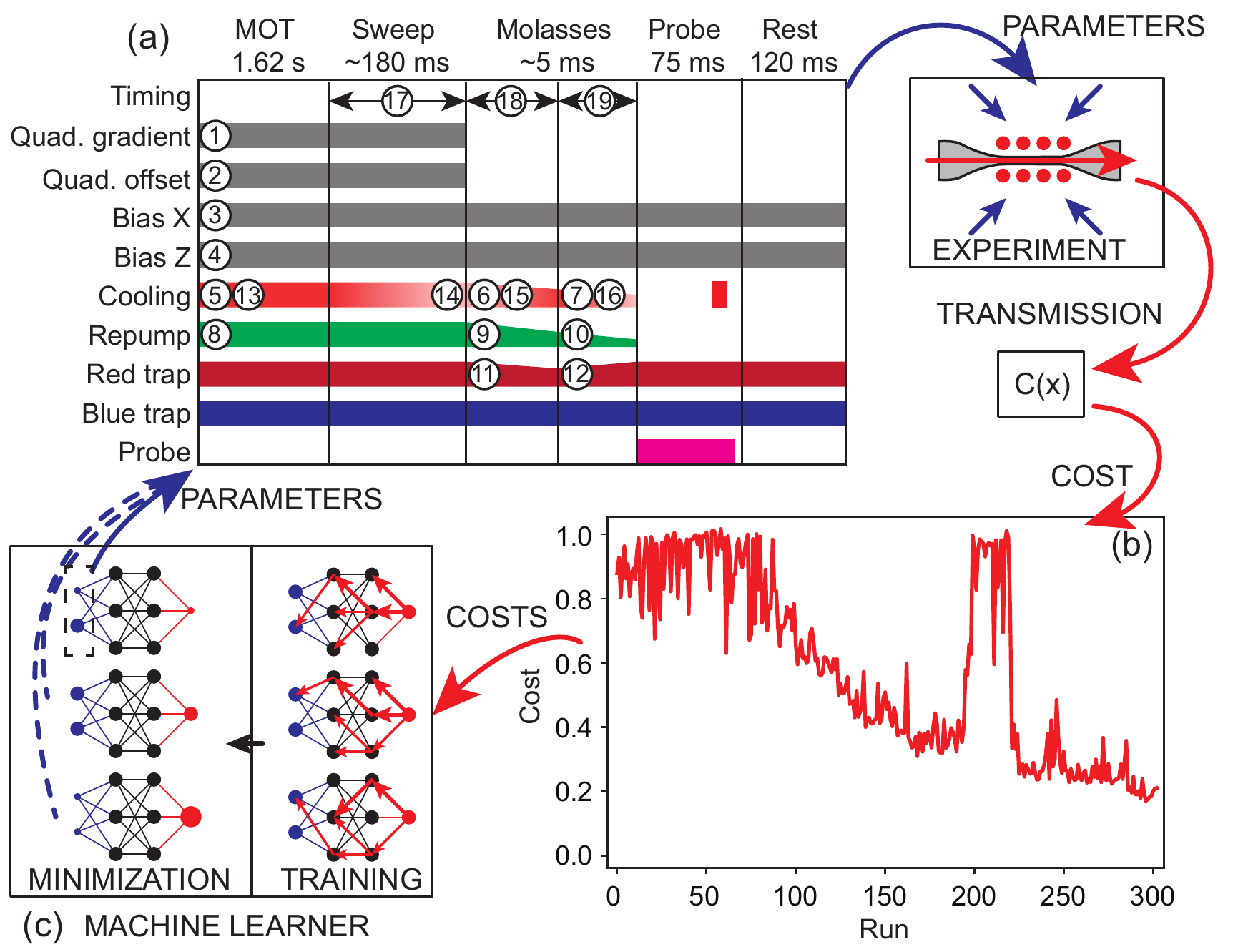}
\caption{Conceptual diagram for online optimization, showing iterative training of the SANN and running of the experiment. A set of parameters (a) are generated by the ML, and the experiment is run with these parameters adjusted. A cost function, $C(x)$, extracts a cost from the experimental data $x$, in this case the averaged transmission of a probe through the ONF, which is appended to the cost-parameter data (b). The data set is used to train the collection of neural networks in the machine learner (c) to map the parameter sets to the corresponding costs. A minimization algorithm is then used to find parameters that give a minimum cost for one of the neural networks (cycling through the networks with each run), and the experiment is run with the new parameters. The entire process is continued for a set number of runs, with the ML predictions improving as the cost-parameter landscape is explored. The timing diagram for the experimental control (a) shows numbered ML-controlled parameters.  1-4: control currents in the MOT magnetic field coils; 5-12: intensities of the 780~nm cooling and repump lasers, and the 1064~nm ONF dipole trap laser at different times; 13-16: cooling laser detunings; 17: duration of the cooling laser detuning sweep; 18-19: durations of the optical molasses stages after the magnetic field gradient is switched off. The cost sequence (b)  shows all the costs after a completed optimization process.}
\label{fig:optimizationflow}
\end{figure}

In this work, we investigated the loading of laser-cooled $^{87}$Rb atoms into an uncompensated ONF-based dipole trap array using machine learning to optimize the final number of trapped atoms. We applied a machine learner (ML) optimization protocol to a control sequence for cooling the atoms in the MOT and subsequently loading them into the dipole trap array.  The ML optimization protocol is based on earlier work \cite{tranter_multiparameter_2018}, in which a predictive agent, in this case a stochastic artificial neural network (SANN), explores a parameter space by predicting new optima based on the results of its previous predictions. The SANN is an ensemble of neural networks acting as surrogate models that constitute the mapping from parameter space to a physical cost; the cost represents the experimental output to be optimized. 'Stochasticity' comes from the independent random initialization of the neural networks in the ensemble. This generates multiple unique representations of the experimental response landscape which are then used to balance the exploration versus exploitation trade-off. Unlike other surrogate methods, the SANN is placed in direct control of the optimization as part of an in-loop feedback. Figure \ref{fig:optimizationflow} illustrates how the SANN, within the ML, controlled the laser cooling experiment and explored the experimental landscape. See Supplement 1 for technical details of the SANN.

\section{Experiment}
\begin{figure}[htbp]
\centering
\includegraphics[width=\linewidth]{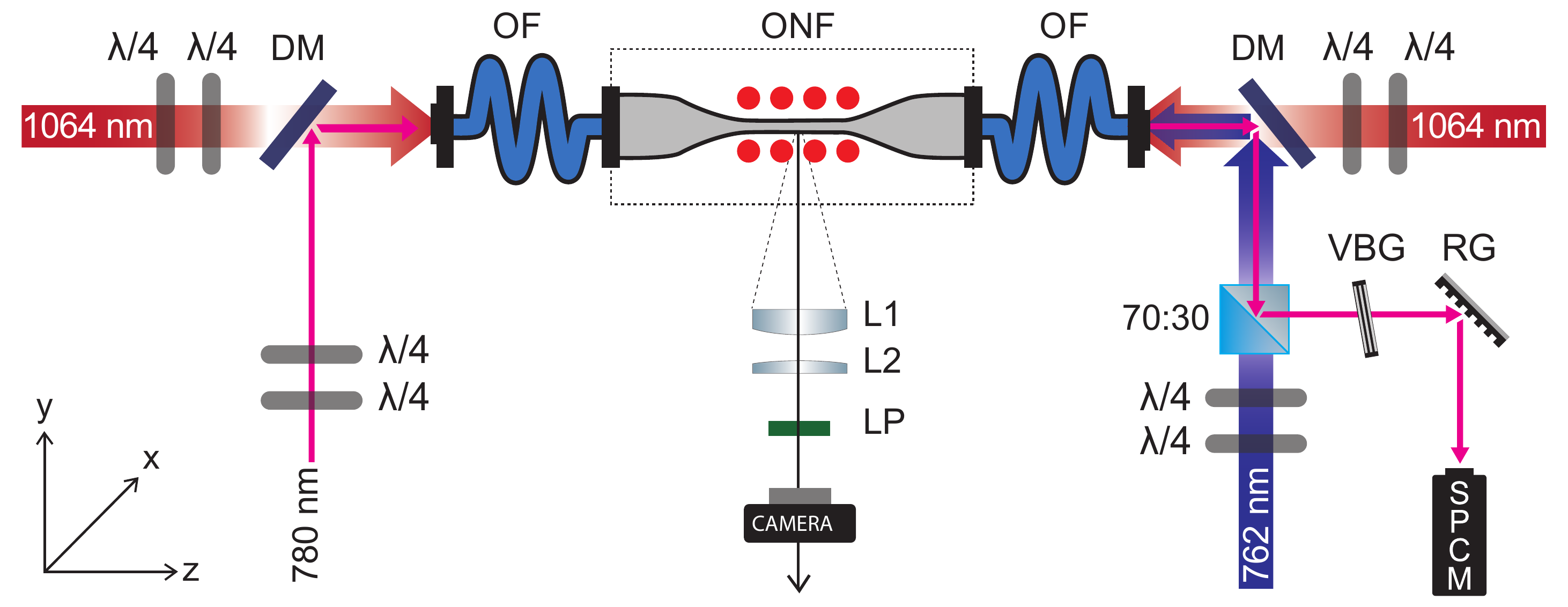}
\caption{Schematic of the experimental setup showing the main optical paths and the ONF-based dipole trap array. DM: dichroic mirror, OF: optical fiber, ONF: optical nanofiber, LP: linear polarizer, VBG: volume Bragg grating, RG: ruled diffraction grating, L1, L2: Lenses, SPCM: single photon counting module.}
\label{fig:setup}
\end{figure}

In the experiments, a $^{\textrm{87}}$Rb MOT was formed around an exponentially tapered optical nanofiber, fabricated from a commercial single-mode optical fiber (Fibercore SM800-5.6-125) \cite{ward2014}, with a $\sim$4 mm long waist region and $\sim$400 nm diameter. The optical dipole trap array was created using a combination of red- and blue-detuned beams relative to the cooling $5S_{1/2}\rightarrow5P_{3/2}$ transition in $^{87}$Rb. We sent 1064~nm light through the ONF in a counterpropagating configuration with 1.8~mW in one direction and 2.1~mW in the opposite direction.  This provided an attractive force for the atoms. Additionally, 1.23~mW of 762 nm light was sent through the ONF along a single  direction to provide a repulsive force.  A weak probe beam (5~pW)  resonant with the $5S_{1/2} (F=2)\rightarrow5P_{3/2} (F'=3)$ transition at 780~nm was sent through the ONF counterpropagating to the 762~nm light. The power of each  beam was measured at its output from the fiber after passing through the ONF.   All the fiber-guided beams were quasi-linearly polarized along the $x$-axis using the method described in \cite{tkachenko_polarisation_2019}. The combined evanescent fields of the fiber-guided light formed a dipole trap array along the waist located at  $\sim200$~nm from the ONF surface. Note that we manually adjusted the polarization of each beam slightly to maximize the probe beam absorption; this had the overall effect of moving the trapping sites closer to the ONF \cite{lee_inhomogeneous_2015}).  A schematic of the experimental setup is shown in Fig.~\ref{fig:setup}.

For both manual and ML optimization, the sequence for cooling atoms in the MOT and then loading them into the dipole traps was computer controlled via a LabVIEW program interfaced with an NI controller PCIe-6363. Timed analog and TTL voltages were used to control acousto-optic modulator (AOM) frequencies and amplitudes controlling the laser frequencies and powers, and to open and close coil circuits in order to switch the magnetic field on and off.  A GPIB controller set the programmable current supplies for the coils.  The experiment took us several months of constant fine tuning to manually optimize. We adjusted the same set of experimental parameters in the same timing sequence as subsequently used for the ML optimization, described below. We also adjusted the cooling and repump beam alignments, which could only be done manually. A typical manual optimization consisted of first maximizing the overlap of the cold atom cloud with the nanofiber during the optical molasses phase by adjusting magnetic field strengths and positions, and the cooling and repump beam alignments. Then, absorption of the probe through the ONF by the dipole-trapped atoms was maximized by adjusting the beam powers and polarizations used for the dipole trap. 

\paragraph{ML optimization}
For the ML-optimized experiments, the above manual optimization process was followed by iterative adjustment of $n=19$ parameters, corresponding to timings and magnetic and optical field powers, via the ML algorithm. This algorithm was running as a separate Python process, see Fig.~\ref{fig:optimizationflow}.

For ML optimization, during an overall 2~s long experimental cycle, the MOT was first operated with constant optical and magnetic fields set by the ML for 1620 ms. The ML controlled the cooling beam detuning (12-20 MHz) and intensity (4-8 mW.cm$^{-2}$), repump intensity, the magnetic quadrupole field gradient (13-15 G/cm), the zero magnetic field position along the $x$-axis (~$\pm$5 mm), and the magnetic field compensation coils (up to 1 G). Numbers in parentheses indicate the range of values the ML was allowed to use. Next, the cooling laser frequency was swept for 180 ms (to 12-36 MHz) before the magnetic gradient coils were switched off. The cooling beam frequency and intensity, and the repump intensity were ramped for 5 ms (divided in two variable length subperiods) to form a 40 $\upmu$K cloud of rubidium atoms. The red-detuned nanofiber dipole trap light intensity was ramped during these 5~ms, but was otherwise kept constant during the experiment in order to prevent thermal stresses from moving the nanofiber. During this final cooling process, cold atoms from the MOT were loaded into the dipole traps.

Once loading was achieved, the cooling and repump lasers were switched off, the probe was detuned $+10$ MHz from the $5S_{1/2} (F=2)\rightarrow5P_{3/2} (F'=3)$ transition and sent through the ONF for 65 ms, and absorption of the probe by the dipole-trapped $^{87}$Rb atoms was measured for the first 10 ms. A dummy cycle was run each time the parameters were changed to equilibrate recapture of atoms in the MOT. The transmission of the probe was integrated over the 10 ms window, normalized, and averaged over the following 3 cycles. The cost function was set as the averaged transmission, $C(x)=T_{probe}$, and was used to train 5 neural networks, see Fig. \ref{fig:optimizationflow}(b). Predictions from the SANN were used for online optimization, according to the method of \cite{tranter_multiparameter_2018} with some minor additions.  The aim was to minimize the cost, and correspondingly increase the absorption of the probe and therefore the number of atoms trapped in the dipole traps. The simple cost function of averaging the probe transmission was used because direct measurement of the atom number after each cycle was too time-consuming. 

After an initial training run of $2n+1$ measured costs of random samples across the allowed experimental parameter space  (where $n$ is the number of parameters being adjusted), the ML ran the experiment with optimum parameters predicted by each of the neural networks in turn, while training all networks with all the existing data after each new cost was returned. When the ML detected convergence (predicted optima continually lying within a small parameter distance), local cost minima outside this region were explored to improve the model around other potential global minima.

On completion of a preset number (usually 300) of trials by the ML, the best measured absorption was tested in a second series of experiments to determine the number of trapped atoms and their average temperature. The transmission of the probe was measured over the entire absorption window of the $5S_{1/2}(F=2)\rightarrow5P_{3/2}(F'=3)$ transition, from -35 to +85 MHz detuning, immediately after the molasses stage, and was typically averaged over 25 experimental cycles. In a separate measurement, a strong probe was also sent to saturate the atoms, with the total absorbed power of the probe used to determine the number of atoms. However, those measurements were not reliable as the force from the probe quickly pushed atoms from the trap.

\paragraph{Modeling of trapped atom spectrum}
To determine the number of atoms loaded into the dipole trap array, the absorption of a fiber-guided probe by the trapped atoms was modeled for various trap powers and temperatures according to the following equation:
\begin{align}
        Abs(\omega)=-N.\sum_{m_F}\int_{Vol_{trap}} OD(m_F,\bf{\mathcal{E}}_{probe}(\bf{r}))\nonumber\\\mathcal{L}(\omega-\delta(\bf{\mathcal{E}_{probe}}(\bf{r}),\it{m}_F))\uprho(V_{opt}(r)+V_{vdW}(r),\it{m}_F,T)d\bf{r}\label{eq:absorption}.
\end{align} 
The atoms were assumed to have a thermal ensemble density $\uprho$ with temperature $T$, due to an approximate van der Waals potential $V_{vdW}$  from the nanofiber surface \cite{daly_nanostructured_2014}, and the optical potential $V_{opt}$ induced by light shifts \cite{le_kien_dynamical_2013,safronova_critically_2011} of the $(5S_{1/2}, F=2, m_F)$ states by the fiber-guided \cite{le_kien_higher-order_2017} trapping fields. The absorption of a probe field $OD(m_F,\bf{\mathcal{E}}_{probe})$ by a transition Lorentzian broadened  $\mathcal{L}$ due to the natural linewidth (ignoring the effect of the fiber on the upper state lifetime) and with a local frequency shift due to the light shifts of the upper and lower levels $\delta(\bf{\mathcal{E}}_{probe}(\bf{r}),\it{m}_F))$, was integrated over the trap volume. The absorption was calculated for the local probe intensity and polarization, with transition strengths according to Clebsch-Gordan (C-G) coefficients and the transition cross section. The atoms had an $m_F$ population due to a simplified optical pumping by the probe field, with $m_F$ density determined by the steady state of a population transfer due to the probe polarization at the trapping potential minimum, with C-G coefficients for the various transitions to  $(5P_{3/2}, F'=3, m_F')$ levels and instantaneous spontaneous decay back to $F=2$ (ignoring stimulated emission and other excitation processes). This model was inspired by that in \cite{lee_inhomogeneous_2015}.

\section{Results}
\begin{figure}[!t]
\centering
\includegraphics[width=0.8\linewidth]{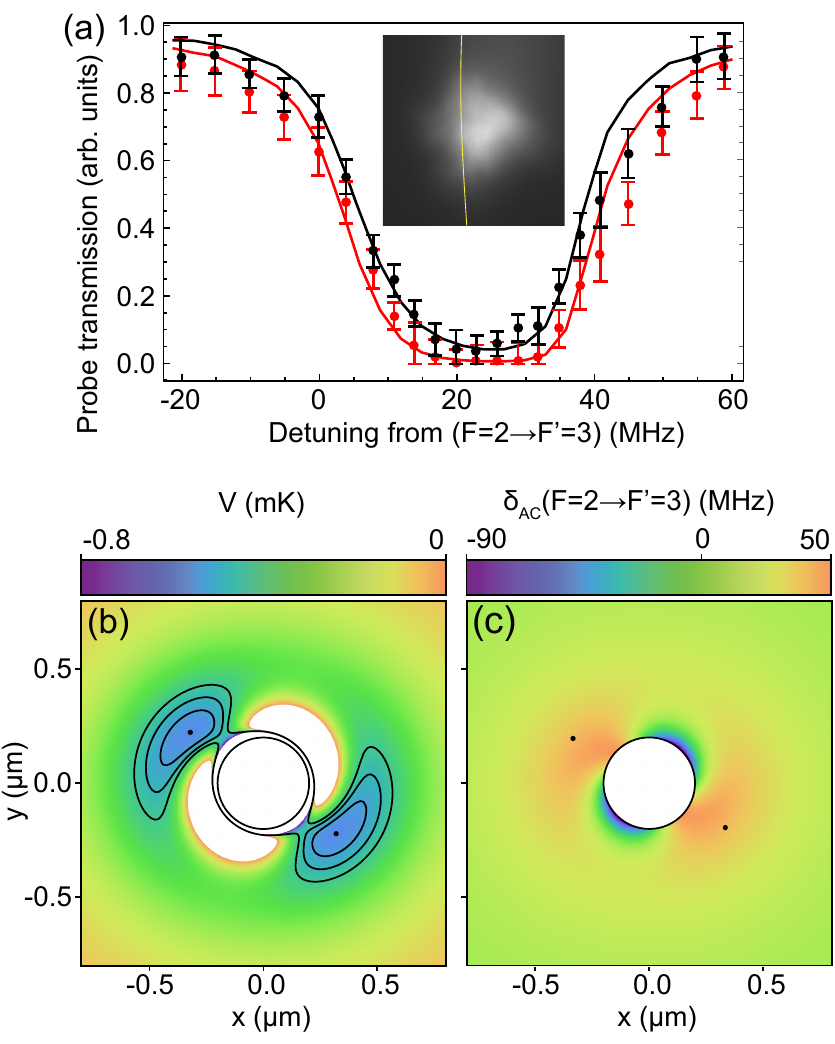}
\caption{(a) Probe transmission spectra after manual optimization (black dots) and ML optimization (red dots) of loading atoms into the dipole traps. Error bars are due to intensity and background noise in the probe measurement. The solid curves are the theory fits for manual optimization (black: 300 atoms at 150 $\upmu$K) and ML optimization (red: 450 atoms at 140 $\upmu$K). Inset: Image of atom cloud at the end of ML-optimized loading, with the ONF highlighted in yellow. (b) Trapping potentials with contours of 50, 100, and 150 $\upmu$K above the minima positions, marked as points. (c) Frequency shift on the probe beam transition  due to the trapping light fields, with the trapping potential minima positions marked as points.}
\label{fig:manualvmachine}
\end{figure}

Figure \ref{fig:manualvmachine} shows the probe beam transmission as a function of detuning for both manually optimized and ML-optimized dipole traps. The theoretical model was fitted using 300 trapped atoms for the manually optimized trap and 450 atoms for the ML-optimized trap. The atoms loaded into the ML-optimized trap were at a slightly lower temperature of 140 $\upmu$K compared to 150 $\upmu$K for the manual trap.  The powers of the dipole trap light fields used to fit the spectra are based on the experimentally measured transmissions at the output of the nanofiber.   Due to losses from the waist to the fiber output, particularly for the 1064~nm light, the actual powers at the waist were higher than those measured.  For example, the measured 1064~nm output power was 2.2 mW and 1.9 mW at either end of the ONF, whereas we used a  modeled power of 3.25 mW in each direction.  For 762~nm light, the output power was 1.23 mW whereas 1.3 mW was used in the model. Note that we had particularly high losses in the ONF likely due to the fact that it was 5 years old and had been used extensively for several earlier experiments \cite{Ray_2020, PhysRevResearch.2.033341, rajasree_generation_2020}.  After the experiment, we removed the fiber from the setup and imaged it using a scanning electron microscope; large  deposits were clearly visible on it and we assume that these were Rb.  These acted as scatterers on the ONF, leading to significant degradation of the transmission through it, for the experiments reported herein. 

The number of atoms trapped in the dipole array was determined from the fit of the modeled spectra, which were calculated for a range of combinations of dipole trap laser powers and atom temperatures, then fit with a skew Gaussian model.  This yielded the central frequency, width, and asymmetry parameters. The experimental data were fit with the same function and the model with the closest matching parameters was chosen. The Gaussian fit width increases with temperature, the central frequency increases with overall trap power, and the asymmetry increases as the trap minima move closer to the fiber. These trends make us confident in the matching of the modeled spectra to the experimental data, clearly indicating an increase in the number of trapped atoms for the ML-optimized system. See Supplement 1 for further details of how the model was used to analyze the experimental data.

\begin{figure}[!htbp]
\centering
\includegraphics[width=0.9\linewidth]{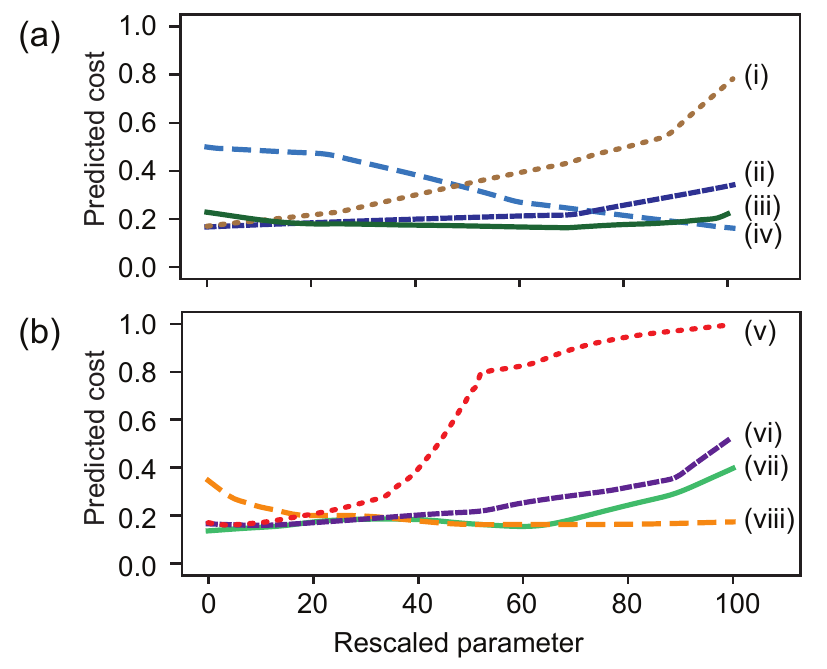}
\caption{Cost predictions from a single neural network after 300 training steps. The cost prediction is the output of the neural network while varying individual optical (a) and magnetic field (b) parameters,  keeping the remaining parameters constant at the predicted optimum. The optical parameters  are those used during the first molasses step. The corresponding parameters are repump laser intensity (i), cooling laser intensity (ii), 1064 nm laser intensity (iii),  cooling laser detuning (iv).  Magnetic field strength of the compensation coils along the $z$-axis (v) and $x$-axis (vi), quadrupole field gradient (vii) and offset in the $z$-normal plane (viii). Parameters are scaled to range from 0 to 100.}
\label{fig:parameterinvestigation}
\end{figure}

Figure \ref{fig:parameterinvestigation} shows the dependence of the learned cost on a selection of the experimental parameters, centered on the best observed. This data is not entirely accurate as the slices represent mostly unsampled areas of parameter space. However, these plots are somewhat useful as they indicate which of the parameters played an important role and which had little role in the ML optimization and can, therefore, be neglected.  They also allow us to determine whether the parameter range should be changed for the ML to explore a different landscape. For example, the repump laser intensity (i), the cooling laser detuning (iv) and the gradient magnetic field offset along the $z$-axis (v) have the strongest effect over their allowed ranges, showing the sensitivity of the cost to these parameters. The numbers in parentheses correspond to the labels in Fig.~\ref{fig:parameterinvestigation}. Due to this sensitivity, we set the magnetic quadrupole parameters as a gradient and position offset to allow the ML to more quickly explore this space (as compared to setting parameters for individual currents through the two quadrupole coils). Learned parameters were stable over time, and reusing the parameters from a single optimization produced a reasonably constant number of trapped atoms over a period of about one month, with no more than $\pm10\%$ change observed. 

\section{Discussion}
The neural network data were typically used to see how effective the training was. For example, whether multiple regions in the parameter space were explored, whether the ML converged on a single region, and whether all the neural networks agreed on the optimum.

The cost function, that is, how the experiment is measured and the data processed to produce the cost, should be carefully chosen. In this experiment the goal was to increase the number of atoms. However, this is measured using transmission as a proxy for experimental simplicity. As a result the optimizer increased the overall optical depth by a larger percentage. The peak optical depth increased by 70\% compared to the 50\% increase in atom number. The increases were sensitive to how long the transmission was integrated for, the delay of the integration from the starting time of the probe, and the optical frequency of the probe. The relationship between optical depth and atom number is not straightforward. To increase the optical depth the ML must balance the temperature, position, and density of the atomic ensemble. These factors, which provide a challenging endeavor for even experienced operators, are implicit in the cost function, thereby allowing the ML to consider them simultaneously.

The ML might also produce specific $m_F$ state populations in the trap, but this would require changes to our experiment so that a specific bias magnetic field could be set after loading the dipole traps.  The probe detection would need to be sensitive enough to measure the absorption prior to optical pumping caused by the probe itself. It could be possible to achieve even further improvements by giving additional parameter controls to the ML.

Relevant to cost function choice is the ability for an optimizer to `cheat' by optimizing for the cost while neither increasing nor even decreasing the number of trapped atoms. We did not observe such behavior, possibly because the ML was never given control of the dipole trap intensities while the probe was sent through the ONF.  Otherwise the ML could shift the spectrum to yield a stronger absorption at the probe detuning.

\section{Conclusion}
We used a machine learner optimizer to increase the number of $^{87}$Rb atoms trapped in ONF-based evanescent field dipole traps from 300 (for manual optimization) to 450.   We derived a microscopic theoretical model that fit the experimental probe transmission spectra and enabled us to determine the number and average temperature of the trapped atoms.  Notably, our manually optimized system showed a very similar number of trapped atoms (300) as that reported in earlier work (302) \cite{lee_inhomogeneous_2015}.  When ML optimized, we increased the number of trapped atoms by 50\% and the optical depth by 70\%.  We expect this to be further improved by using a new ONF that does not show such high transmission losses.   Additional investigations of the capabilities of this setup are planned, including optimizing the loading of atoms for a collective atom-light interaction such as four-wave mixing. We envision that this will be feasible with the obtained number of atoms.  The techniques developed herein can also be extended to any atomic species where it is desirable to increase the atom number.

\begin{backmatter}
\bmsection{Funding} This work was supported by OIST Graduate University and JSPS Grant-in-Aid for Scientific Research (C) Grant Number 19K05316.

\bmsection{Acknowledgments} This work was supported by OIST Graduate University.

\bmsection{Disclosures} The authors declare no conflicts of interest.

\bmsection{Data availability} Data underlying the results presented in this paper are not publicly available at this time but may be obtained from the authors upon reasonable request.

\bmsection{Supplemental document}
See Supplement 1 for supporting content. 

\end{backmatter}

\bibliography{bib}

\clearpage

\section{Supplemental Material}
\subsection{Theory}
The trapping potential of an atom in an optical dipole trap-induced dipole potential, and the consequent broadening of the transition energy, is due to the ac-Stark shift of the atomic energy levels. The potential is given by the polarizability, $\alpha$, and the optical intensity, $|\mathcal{E}|^2$, as

\begin{equation}
    V = -\frac{1}{2}\alpha|\mathcal{E}|^2\label{eq:potential},
\end{equation}

where  $\alpha$ can be decomposed as the sum of scalar ($\alpha^0$), vector ($\alpha^1$), and tensor ($\alpha^2$) polarizabilities:

\begin{align}
    \alpha(\omega) = \alpha^0(\omega) &+ \mathcal{A} \cos{\theta_k} \frac{m_F}{F} \alpha^1(\omega)\nonumber\\
    &+ \bigg\{\frac{3\cos^2{\theta_p}-1}{2}\bigg\}\frac{3{m^2_F}-F(F+1)}{F(2F-1)}\alpha^2(\omega)
    \label{eq:polarizability}
\end{align}

where $\mathcal{A}$, $\theta_k$ and $\theta_p$ describes the degree of circular polarization, the angle between the wave vector and the $z$-axis, and the angle between the polarization and the $z$-axis, respectively. The dynamic polarizability depends on the detuning of the light from the various dipole transitions available to the atom.

Importantly, the presence of a longitudinal electric field component in the evanescent field of the ONF, with a $\pi/2$ phase relative to the transverse components causes the field to have non-zero ellipticity. This is a crucial distinction between optical dipole traps using free-space beams and those using the evanescent field. This gives stricter requirements on 'magic' wavelengths in evanescent ONF dipole traps, including the requirement to choose particular $m_F$ levels to compensate. The effect of the longitudinal field is reduced in the standing wave, which is used for red, and the traveling wave, which we do not use for blue.

It is straightforward to calculate the evanescent fields and resulting shifts of the atomic energy levels. However, the frequency dependence of the atomic absorption in an uncompensated trap is complicated enough that a physical model is necessary to interpret the experimental results.

\subsection{Modeling the absorption spectra}
Modeling was done by calculating the absorption due to the trapped atoms, for a range of frequencies to compare to the experimental data. Because the experimental trap intensities and polarizations were only approximately known, a range of these parameters were tested, and trends were found to assist with matching to the experimental data.

The integral in the main text is carried out only over one 'lobe' of the trap to avoid having to use different $m_F$ occupations for the other lobe, where the polarization is different but the absorption asymmetry is the same due to the reversal of the $m_F$ occupation. The integral is performed with a Monte Carlo sampling method in Mathematica, which is significantly faster than a grid recursion method, for a small sacrifice of precision.

Modeling is done with a variety of quantization axes. The inclusion of optical pumping by the probe greatly reduces the effect on the modeled optical depth due to the choice of quantization axis. A quantization axis in the $x$- (optical polarization axis) or $y$-axis, i.e., perpendicular to the $z$-axis, which is the fiber axis, gives a fit that predicts roughly the same atom number.

Effects neglected in the model include the change in the linewidth of the atoms due to the presence of the fiber, rescattering into the fiber, motion of the atoms during absorption, and detuning of the probe and individual $m_F$ levels within the optical pumping scheme.

\begin{figure}
    \centering
    \includegraphics[width=0.8\linewidth]{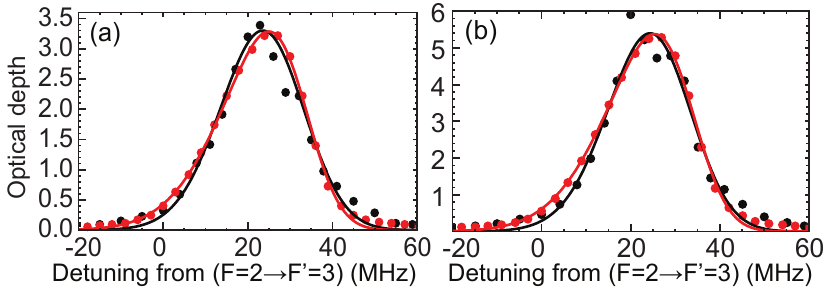}
    \caption{(a) Shows optical depth versus probe detuning for the manually optimized trap, while (b) shows the same for the spectrum of the ML optimized trap. Black and red points show the experimental and modeled optical depths, respectively. Red and black lines are the fitted functions to the data points.}
    \label{fig:suppfig1}
\end{figure}

\paragraph{Trends for fitting}
Absorption spectra of the dipole trapped atoms were modeled for a range of different trap powers and atom temperatures, and then the spectrum that most closely matched the experimental result was selected. The match was quantified by fitting the modeled and experimental spectra with a skewed Gaussian function, and then choosing the modeled spectrum whose fitted asymmetry, central frequency, and broadening parameters most closely matched the experimental spectrum fitting. Figure \ref{fig:suppfig1} shows the fitted functions for the manual and ML optimized trap spectra.

The skew gaussian fits are better for the model data than the experimental data. In the experimental data, the higher absorption at around 50 MHz detuning is possibly due to atoms that are trapped in tight orbits around the fiber by a combination of van der Waals forces and repulsive blue-detuned light. The positive light shifts close to the fiber max out around 50 MHz, and even a few atoms could absorb this much probe light. However, it was not possible to include these atoms in the model. There is not much difference in the 50 MHz absorption in the two experimental spectra, so we ignore their effect on the fits.

\subsection{Stochastic Artificial Neural Network}

The stochastic artificial neural network (SANN) consists of an ensemble of identical neural networks which are trained on identical data but initialized independently. Each neural network provides a mapping between parameter space and cost space representing an $n$ to 1 mapping, where $n$ is the number of controllable parameters. As this is a form of supervised learning, training data takes the form $\{ \boldsymbol{X}, \boldsymbol{c} \}$ where $\boldsymbol{X}$ is a $m\times n$ matrix representing $m$ inputs, and $\boldsymbol{c}$ is a vector of $m$ associated costs. 

The networks within the SANN have an identical structure characterized by an input layer of dimension $n$, followed by 5 densely connected layers. Each dense layer has 64 neurons per layer with Gaussian Error Linear Unit (GELU) activation \cite{hendrycks2016gaussian}. L2 regularization \cite{krogh1992simple} is also applied to moderate overfitting behavior. The networks are trained by minimizing the mean squared error using the Adam optimizer \cite{kingma2014adam}. Early stopping is also used to mitigate overfitting.

An initial sampling policy is used to build a training set to learn an initial representation of the cost landscape. After this initial training the SANN prediction loop is instantiated. Each network in the SANN generates a set of test parameters by performing a minimization search of the surrogate landscape using parallel L-BFGS instances \cite{byrd1995limited}. Each prediction is tested experimentally with the resulting cost being added to the training data. The networks within the SANN are trained and minimized asynchronously to reduce the time taken to generate predictions. This feedback loop continues until some stopping criteria is met. For non-convex optimization problems the idea of convergence is ill-defined without prior knowledge of the problem.

\end{document}